\documentclass[aps,pre,twocolumn,groupedaddress, showpacs]{revtex4-1}
\usepackage{epsfig,amssymb,amsmath,graphicx,subfigure,hyperref}
\usepackage{tabularx}
\usepackage{float}
\usepackage{setspace}
\usepackage{color}

\usepackage{array}

\begin{document}

\title{Planar and curved droplet networks}

\author{Duanduan Wan and Mark J.~Bowick}
\email[Email: ]{bowick@phy.syr.edu}
\affiliation{Department of Physics, Syracuse University, Syracuse NY 13244, USA}

\date{\today}

\begin{abstract}
We study how a two-dimensional square sheet of droplets evolves to the preferred triangular lattice. The rearrangement can be induced either by an impurity seed for flat sheets or by curving the sheet. We analyze the propagation of transformation fronts and the spatial reordering subsequent to the initial perturbation.
\end{abstract}

\maketitle

\section{Introduction}
\label{intro}

Metastable states, local minima of the free energy landscape, are very common. Transitions from one local minimum to another local minimum (not necessarily the ground state) may take place spontaneously, or when system parameters, such as the ambient temperature and pressure, are changed. They may also be triggered by external perturbations or the introduction of impurities. Examples include avalanches \cite{Tsonis2007}, protein folding \cite{Alberts2002}, devitrification of glass \cite{Vogel1994}, crystallization of supercooled liquids \cite{Moore2011} and aggregation and sedimentation of colloidal systems \cite{Russel1992}. One broad class of such transitions is solid to solid phase transitions. Diamond, for example, is not thermodynamically stable at standard temperature and pressure {--} it exists because of the large activation barrier for conversion to graphite \cite{Bundy1980}. Martensite, which results from rapid cooling, is also a metastable phase and is easily destroyed by heating \cite{Porter1992}. Despite their physical significance the detailed dynamics of solid to solid phase transitions is difficult to observe experimentally as it is rapid and typically occurs on small length scales.  Colloidal crystals, whose dynamics are significantly slower, provide a rich analogue system to their atomic or molecular counterparts \cite{Ise2005,Gasser2009,Zhou2011}.

An entirely different model system is that of droplet networks \cite{Bayley2013} formed  by printing tens of thousands of micron-sized aqueous droplets that bind by forming single lipid bilayers \cite{Princen1980,Bayley2011} to form a cohesive network in bulk oil.  Functionality can be added to the network and the droplets can be programmed with different osmolarities (osmolarity is the measure of solute concentration) to fold the network into certain designed structures \cite{Bayley2013,Tao2015}. Here instead of discussing the promising applications of the network, we will explore the potential possibilities of using the droplet network as a platform to study lattice transitions. Both the colloidal crystals and the droplet systems are platforms to study lattice transitions: Colloidal crystals have long been studied, while the droplet systems have their own advantages such as well-defined interaction \cite{Bayley2013}, programmable initial states \cite{Bayley2013} and reversibility of a folded shape \cite{Tao2015}. As the initial positions of the droplets are predetermined in this system, the droplets can be arranged in metastable states. This is an ideal system to study lattice transitions. The interaction between droplets may be simply modeled as a spring potential with a damping term \cite{Bayley2013}. We examine a simple metastable state -- a two-dimensional (2D) square network -- and study its transformation into a triangular lattice. This transformation has also been investigated in a study of temperature-induced lattice transitions in colloidal crystals \cite{Han2015}. Osmolarity gradients lead to flow of water from lower to higher osmolarity, resulting in the swelling of the higher osmolarity droplets and the shrinking of the lower osmolarity droplets. We thus use osmolarity gradients to introduce size polydispersity. The effects of size polydispersity have been analyzed for a variety of physical processes, such as crystallization \cite{Lekkerkerker2005} and granular dynamics \cite{Menon2002}. Yao et al. \cite{Yao2014} suggest that polydispersity-driven topological defects could act as order restoring excitations. We start by considering the situation where there is an impurity of larger size in the square lattice. Surrounded by smaller particles, the coordination number of the impurity tends to exceed four, destroying the regular square lattice and initiating the rearrangement. We focus on the dynamics of the transformation, rather than minimal energy states \cite{Yao2014,Thomson1904,Yao2013,Funkhouser2013}, with the final configurations not necessarily the ground states. After the  rearrangement is initiated, we explore how the subsequent propagation of reordering relates to the strengths of the interaction and the damping. For two initial impurities we find that the competition between transformation fronts, as the reordering propagates, leads to the formation of an intervening domain wall.

We also consider the motion of droplets on curved surfaces, motivated by previous studies of colloidal particles confined on a two-sphere \cite{Bausch2003} as well as on capillary bridges \cite{Irvine2010,Irvine2012,Bowick2011}, and explore curvature-induced transformations of the lattice configuration of the droplet network. This may be realized experimentally by choosing appropriate hydrophilic substrates. In the quasi-static process of slowly increasing the curvature of an elliptic paraboloidal surface we observe a first transition after which only the boundary regions transform into the triangular lattice structure and a second transition after which a cross shape forms in the central area. Rearrangements take place in stages  also for the case of a paraboloidal surface. For impurity-driven transformations on a spiral surface we find that ordered domains are rotated in the final configuration.

\section{Model and Methods}
We briefly describe the model of Ref.~\cite{Bayley2013}. Lipid-coated aqueous droplets in a continuous oil phase adhere at their interfaces by forming stable bilayers. To simulate this mechanical interaction, every droplet $i$ is treated as a point mass with an associated radius $R_{i}$. If two droplets $i$ and $j$ come within a distance $R_{i}+R_{j}$, they become connected by a Hookean spring of equilibrium length $\alpha(R_{i}+R_{j})$: the spring represents the attractive interaction between fused droplets and the parameter $\alpha$, which is less than 1, approximates droplet deformation due to droplet-droplet adhesion. The net force on each droplet $i$ is therefore

\begin{equation}
\vec{F}_{i} = \sum_{j\in C_{i}}k \left[ r_{ij}-\alpha \left(  R_{i}+R_{j} \right) \right] \hat{\vec{r}}_{ij} -\gamma \vec{v}_{i}, 
\label{Force}
\end{equation} 
where $C_{i}$ is the set of droplets connected to the $i$th droplet, $\vec{r}_{ij}$ is the vector from droplet $i$ to droplet $j$ with magnitude $r_{ij}$, $k$ is a spring constant, $\gamma$ is a damping coefficient, and $v_{i}$ is the velocity of the $i$th droplet. The force increases linearly with the distance between two droplets until the distance reaches  $R_{i}+R_{j}$, where the force drops to zero and remains zero for distance above. The position of each droplet is updated using 

\begin{equation}
\vec{F}_{i} = m \frac{d^{2} \vec{r}_{i}}{dt^{2}},
\label{Newton}
\end{equation} 
where $m$ is the mass of the droplet. The equation of motion for a droplet is given by

\begin{equation}
\frac{d^{2} \vec{r}_{i}}{dt^{2}} = \sum_{j\in C_{i}} \omega^{2} \left[ r_{ij}-\alpha \left(  R_{i}+R_{j} \right) \right] \hat{\vec{r}}_{ij} - \omega \zeta \vec{v}_{i},
\label{Force}
\end{equation}
where $\omega =\sqrt{k/m}$ is the angular frequency and $\zeta=\gamma/\sqrt{mk}$ is the dimensionless damping coefficient.

\section{Results and Discussion}
\subsection{Impurity-induced transformation}

Defects play an essential role in understanding 2D crystalline systems. Defects on a shell, for example, may drive the shape from round to faceted, depending on the F{\" o}ppl-von K\'{a}rm\'{a}n number \cite{Lidmar2003}, and the precise defect distribution affects the shell's ability to sustain external pressure \cite{Wan2015}. Here we use osmolarity gradients to introduce impurities, and analyze how the subsequent propagation of rearrangements depends on the values of the spring constant $k$ and the damping parameter $\zeta$. Osmosis is implemented as follows: when two droplets with different osmolarities are connected by a spring, the volume of water transferred per unit time from a droplet $i$ with osmolarity $C_{i}$ to a droplet $j$ with osmolarity $C_{j}$ is given by Fick's first law as

\begin{equation}
J_{ij}=A_{ij}D\left(C_{j}-C_{i} \right),
\label{Exchange_water}
\end{equation}
where $A_{ij}$ is the area of the bilayer between the two droplets, and $D$ denotes a permeability coefficient which is assumed to be constant and identical for all bilayers. 
Following Ref.~\cite{Bayley2013}, we assume $k$, $m$ and $D$ are the same for all droplets.

The positions of the droplets are updated in the following way: at time $t$, any two droplets $i$ and $j$ that have come within a distance $R_{i}+R_{j}$ are connected by a spring of natural length $\alpha(R_{i}+R_{j})$. If the osmotic interaction is active, the volume of water transferred between each pair of droplets joined by a bilayer is calculated according to Eq.~(4), and the size of each droplet is updated accordingly. The position of each droplet at time $t+\triangle t$ is calculated according to Eqs.~(1) and (2) using a fourth-order Runge-Kutta scheme. All the droplets have uniform initial volume $V_{0}= 4\pi R_{0}^{3}/3$, where $R_{0}$ is the initial radius. We set $R_{0}=1$, $m=1$, $\alpha=0.8$, $D=1$ ($D$ is chosen larger than that in Ref.~\cite{Bayley2013} to shorten the time required for the initial water exchange - this has little effect on the subsequent propagation of rearrangements), $\triangle t= 10^{-3}$ (in the large damping case Fig.~1f, $\triangle t= 2\times 10^{-4}$ to make the iteration stable). The units of length, mass, spring constant and osmolarity can be obtained by matching their experimental values given in Ref.~[11]; the units of other quantities are a combination of these units.

\begin{figure}
\centering 
\includegraphics[width=3.45in]{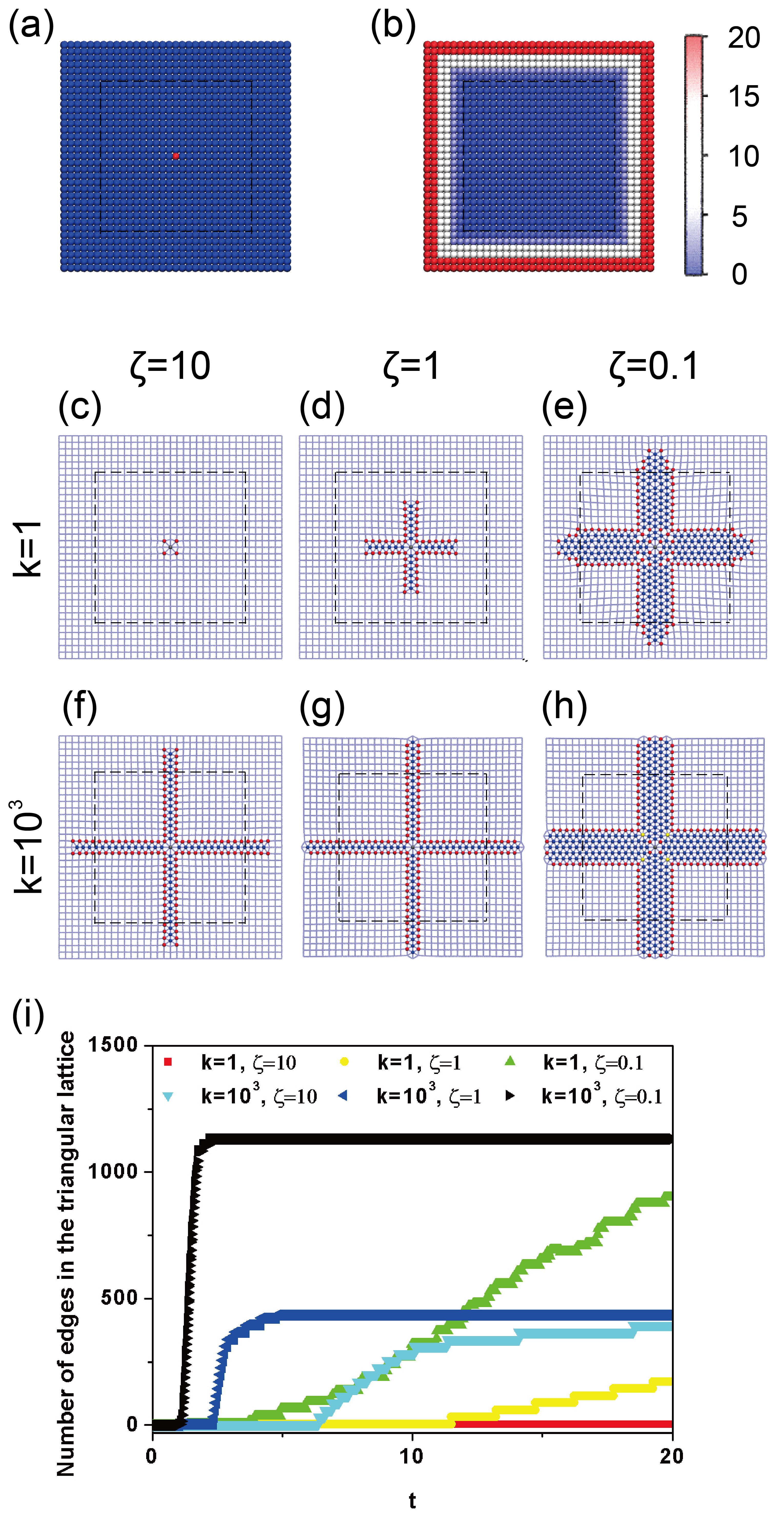} 
\caption{(Color online) (a) Initial osmoloarity profile of a square lattice in a square area, composed of $N=1225= (N_{L}+N_{B})\times(N_{W}+N_{B})=(23+12)\times(23+12)$ droplets, with $N_L$, $N_{W}$, $N_{B}$ being the number of droplets along the length, along the width, in the boundary (outside the black dashed line), respectively. An impurity of high osmolarity $5/V_{0}$ (red) at center and others of uniform low osmolarity $1/V_{0}$ (blue). (b) The absorptive boundary condition in the case $\zeta=1$. Color represents $\zeta$ value. Internal  droplets are with $\zeta=1$. From innermost to outermost, $\zeta= 2, 4, 10, 10, 20, 20$, respectively, for the six layers in the boundary. (c-h) Connectivity at $t=20$. Fivefold vertices are shown in red, sixfold in blue, sevenfold in yellow and eightfold in grey.  
(c-e) Fixed $k=1$, $\zeta=10, 1, 0.1$. 
(f-h) Fixed $k=10^{3}$, $\zeta=10, 1, 0.1$.
(i) Number of edges with triangular order for various $k$ and $\zeta$ values. Snapshots were generated using the Visual Molecular Dynamics (VMD) package \cite{Humphrey1996} and rendered using the Tachyon ray tracer \cite{Stone1998}.}  
\label{square}
\end{figure}

We start by studying the situation of one impurity. Figure~\ref{square}a shows the initial configuration of $N=1225$ droplets positioned in a square lattice, with one impurity of high osmolarity $5/V_{0}$ at the center and all others of uniform low osmolarity $1/V_{0}$. These droplets are first equilibrated without water exchange. To minimize boundary effects we implement absorptive boundary conditions by gradually increasing the $\zeta$ values of the outermost six layers. The setup is illustrated in Fig.~\ref{square}b: the internal droplets have $\zeta=1$ the six outermost layers have $\zeta=2, 4, 10, 10, 20, 20$ in order of innermost to outermost. With these $\zeta$ values we do not observe any reflective propagations. The boundary conditions when the internal droplets have other $\zeta$ values are given in Appendix A. To better describe the system, we define the connectivity of the system as the structure of the spring network at a given time. After water exchange is turned on the impurity swells and connects with the four next-to-nearest neighbors {--} the coordination number of the impurity seed thus doubles from four to eight. The four next-to-nearest neighbors acquire coordination number five. This initiates the propagation of a cross-shaped region of triangular lattice within the original square lattice. Away from the impurity, the additional kinetic energy comes from the potential energy of the initial system through the formation of extra bonds and the propagation speed depends on the stiffness ($k$) and the damping ($\zeta$). Either higher stiffness or lower damping $\zeta$ increase the conversion rate to the triangular lattice. We plot the total number of edges of the converted triangular lattice, for a variety of stiffnesses and damping values, versus time in Fig.~\ref{square}i and the corresponding connectivities at $t=20$ in Fig.~\ref{square} (c-h). In our simulation time the rearrangement is localized in the case of small stiffness and high damping (Fig.~\ref{square}c and movie \ref{square}c). For higher stiffness or lower damping the coarsening transformation front propagates (Fig.~\ref{square}, d-h and movie \ref{square}, d-h). The propagation speed is almost constant, as can be seen in Fig.~\ref{square}i, until the front reaches the boundary region where it slows down. Figure~\ref{square}d will propagate to the boundary provided longer simulation time. Triangular domains and square domains are separated by lines of fivefold defects. The position of the defect lines may depend on $k$, $\zeta$ and the initial perturbation.

We note that both $k$ and $\zeta$ may depend on the temperature, and the effect of temperature can be included as we can change $k$ and $\zeta$ for different temperatures. In the following we choose $k=10^{3}$ and $\zeta=0.2$, similar to the values used in Ref.~\cite{Bayley2013}. We adopt absorptive boundary conditions for impurity-induced transformations. 

\begin{figure}
\centering 
\includegraphics[width=3.2in]{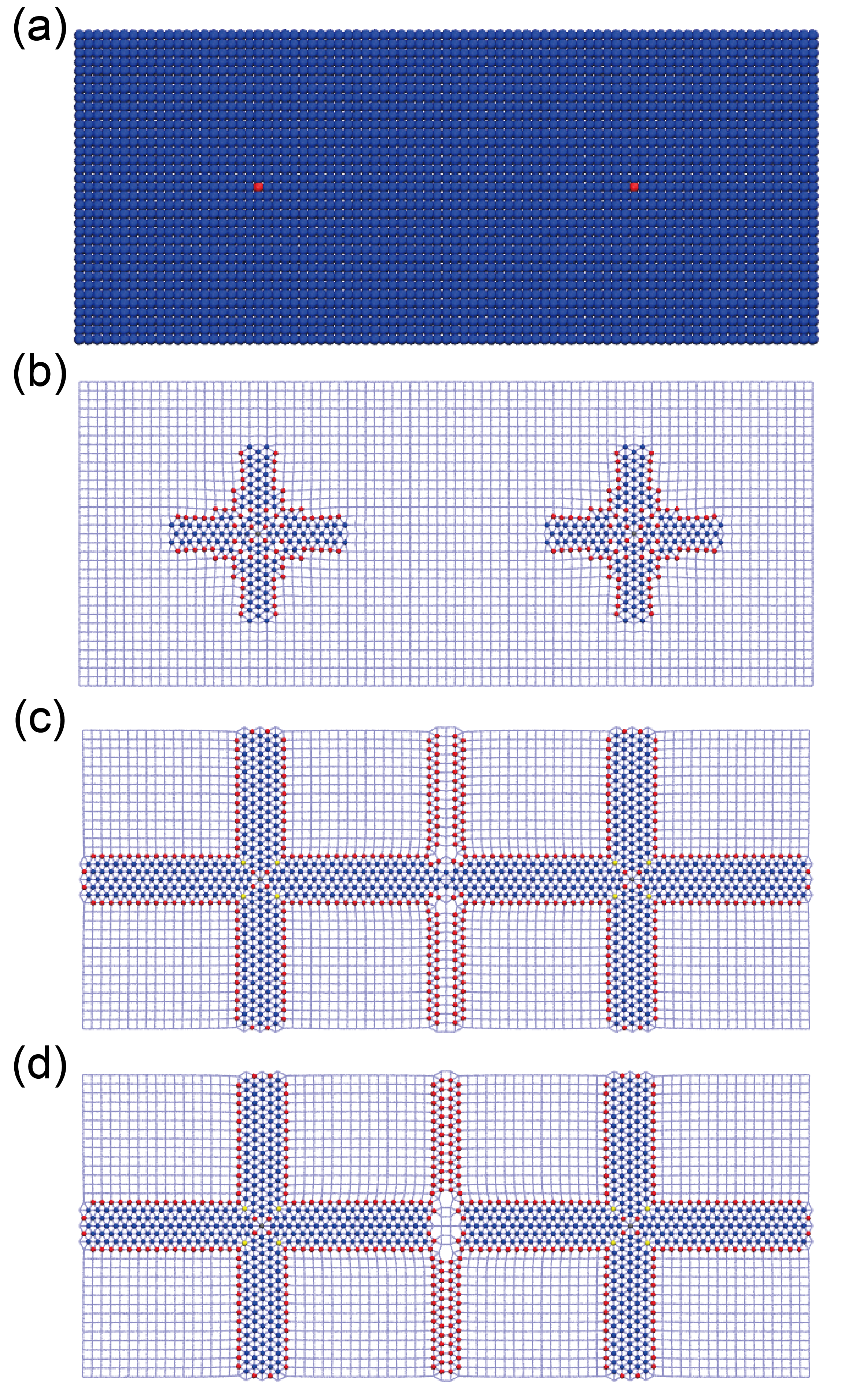} 
\caption{(Color online) (a) Initial configuration of $N=2905$ ($N_{L}=83$, $N_{W}=35$, boundary not included) droplets with two impurities of high osmolarity. The number of columns between the pair of impurities is $n_{p}=41$. Colors represent osmolarity, as in Fig.~1a. (b) Connectivity at $t=1.4$. (c) Connectivity at $t=10$. (d) Connectivity at $t=10$ for a $N_{L}=82$, $N_{W}=35$, $n_{p}=40$ system.} 
\label{rectangular}
\end{figure}

We treat now the case of two impurities. Figure~\ref{rectangular}a shows the initial  configuration {--} a square lattice of $N=2905$ droplets with a pair of impurities of high osmolarity. The distance between the pair of impurities is 42 lattice spacings. Figure~\ref{rectangular}b shows the connectivity of this system at time $t=10$. Coarsening fronts continue to propagate to the boundary after meeting in the region between the impurities, leading to the formation of a domain wall with vacancies (Fig.~\ref{rectangular}, c and d; movie \ref{rectangular}, c and d). Different final states could be obtained if we arrange the positions of the two impurities differently or if we consider more impurities.

\subsection{Effects of randomness}
\begin{figure}
\centering 
\includegraphics[width=3.4in]{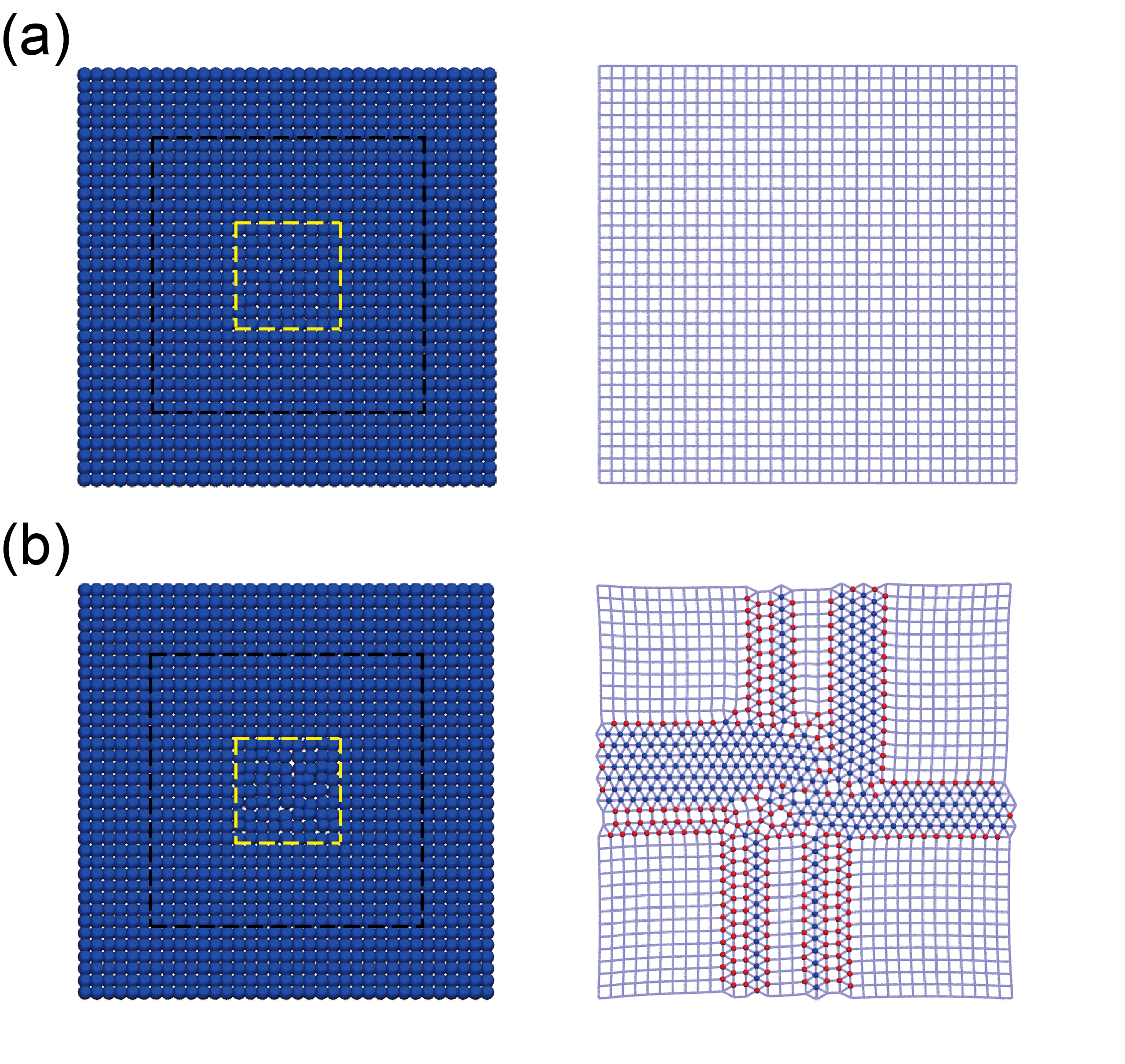} 
\caption{ (Color online) Left: Initial configurations with small perturbation $d=0.13$ (a) and larger perturbation $d=0.3$ (b) in the central regions (the yellow dashed line). The perturbation can be seen from the irregular white space between droplets. The black dashed line indicates the boundary. Right: Connectivity at $t=10$ after relaxation.}  
\label{random}
\end{figure}

As there may be stochastic noise in the experiment, we study the effects of the randomness of the droplet positions. In Fig.~\ref{random},  we give the droplets in the central region (inside the yellow dashed line) a perturbation in a random direction, with a maximal value $d$, as the initial state. Then we relax the system. When $d$ is small, the connectivity of the square lattice is preserved. After relaxation, the droplets tend to return to their equilibrium positions, as shown in Fig.~\ref{random}a(right), where we see a nearly perfect square lattice connectivity. Note that the final state may be slightly different from a perfect square lattice even after running the simulation for a rather long time (say $t \approx 50T$, with $T$ the period and given by $T=2\pi / \omega$) due to the damping term. On the contrary, when $d$ is large enough, the initial square lattice will break down without introducing additional defects. After relaxation, the transformation from the square lattice to the triangular lattice will spread out of the core region and reach the boundary, as shown in  Fig.~\ref{random}b(right). That the effects of randomness depend on strength is understandable as the square lattice is a metastable state and thus has the ability to sustain small perturbations. Provided the noise in an experimental setting is not large and the initial positions of the droplets are carefully arranged, and subsequently allowed to relax for a  sufficiently long time, we expect our simulation results to provide an accurate guide.

\subsection{Curvature-induced transformation}
Previous studies have explored chemically-driven free-running droplets on solid substrates \cite{Santos1995,Lee2000,Sumino2005,John2005}. Here we shift to treating the case of droplets constrained to move on curved surfaces which can arise due to osmolarity-gradient induced bending of an initially flat surface. For simplicity we assume that droplets on the surface are not subject to any pressure (we don't consider the motion in the normal direction). With this assumption, it can be shown that the the differential equations of motion of a droplet of mass $m$, constrained to move in a smooth surface $S$: $\mathbf{x}\left( u,v \right)$, read \cite{Decicco1947} 
\begin{eqnarray}
\ddot{u} = \phi - \Gamma_{11}^{1} \dot{u}^{2} - 2 \Gamma_{12}^{1} \dot{u}\dot{v} - \Gamma_{22}^{1} \dot{v}^{2} 
\end{eqnarray}
and 
\begin{eqnarray}
\ddot{v} = \psi - \Gamma_{11}^{2} \dot{u}^{2} - 2 \Gamma_{12}^{2} \dot{u}\dot{v} - \Gamma_{22}^{2} \dot{v}^{2},
\end{eqnarray}
where $\Gamma_{\alpha \beta}^{\gamma}$ $(\alpha, \beta, \gamma=1,2)$ are Christoffel symbols \cite{Kreyszig1991}, $\phi$ and $\psi$ can be written in term of the force vector $\vec{F}$ defined in the Euclidean space and given by Eq.~(1) as $ \phi =  [g_{22}(\vec{F}\cdot \partial \mathbf{x}/\partial u)- g_{12}(\vec{F}\cdot \partial \mathbf{x}/ \partial v)]/mg$ and $ \psi =  [-g_{12}(\vec{F}\cdot \partial \mathbf{x}/\partial u )+ g_{11}(\vec{F}\cdot \partial \mathbf{x}/ \partial v)]/mg$, with $g_{\alpha \beta}$ $(\alpha, \beta =1,2)$ the metric tensor and $g = \mathtt{det} (g_{\alpha \beta}) =g_{11}g_{22}-(g_{12})^{2}$. The interaction between droplets and the damping effect are included in the $\phi$ and $\psi$ terms; the curvature effect is included in the other three terms which vanish in flat geometry in each equation.

\begin{figure}
\centering 
\includegraphics[width=3.2in]{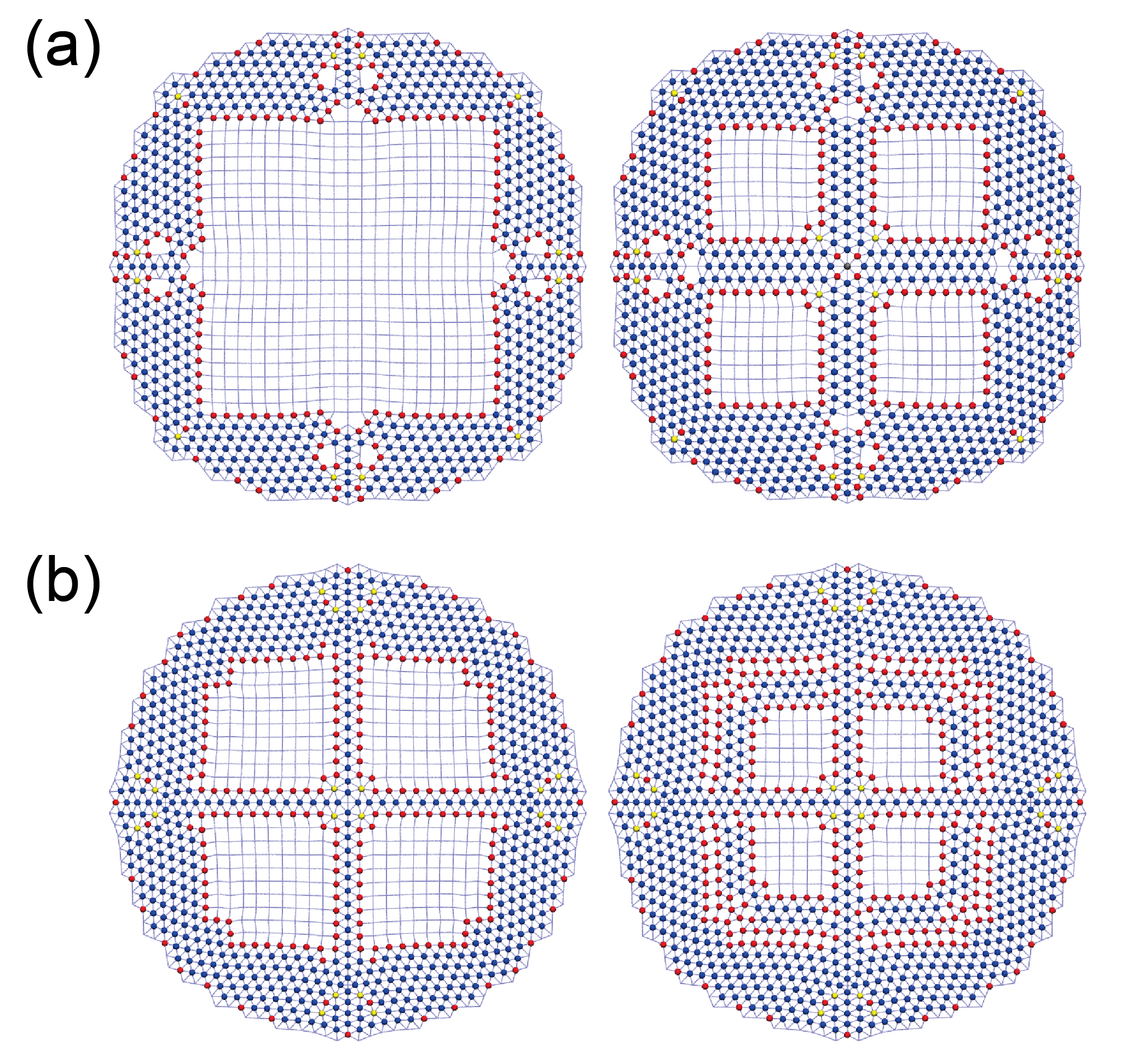} 
\caption{(Color online) $N=1245$ droplets constrained to move on an elliptic paraboloid surface (a) and a hyperbolic paraboloid surface (b). (a) Connectivities after the transitions at $a\approx8.9$ (left) and at $a\approx7.2$ (right). (b) Connectivities after the transitions at $a\approx8.9$ (left) and $a\approx 8.4$ (right).} 
\label{curvature}
\end{figure}

We start by studying a curvature-induced transformation. $N=1245$ droplets of uniform osmolarity are positioned within a circular area $\sqrt{x^{2}+y^{2}} < 20$ on an elliptic-paraboloid surface given by $z=x^{2}/a^{2}+y^{2}/a^{2}$, with $a=15$, and subsequently equilibrated. For this elliptic-paraboloid surface, $\Gamma_{11}^{1}=\Gamma_{22}^{1}=4x/(a^{4}+4x^{2}+4y^{2})$, $\Gamma_{11}^{2}=\Gamma_{22}^{2}=4y/(a^{4}+4x^{2}+4y^{2})$, $\Gamma_{12}^{1}=\Gamma_{12}^{2}=0$. To simulate a quasi-static process we relax the system for $2\times10^{3}$ time steps every time $a$ is decreased by $0.1$. We observe a first transition (movie \ref{curvature}a) at $a\approx8.9$, with boundary regions transforming into a triangular lattice with defects, while the central region preserves the square lattice structure. The connectivity after this transition is shown in Fig.~\ref{curvature}a (left). As $a$ further decreases, we find a second transition happens at $a\approx7.2$, with the formation of a cross shape in the central region (Fig.~\ref{curvature}a, right). After this transition the remaining square lattice gradually transforms into triangular lattice as $a$ decreases. The situation is much the same for droplets confined on a hyperbolic paraboloid surface, $z=x^{2}/a^{2}-y^{2}/a^{2}$ with $a=15$.
We observe two transitions (movie \ref{curvature}b), the first at $a\approx8.9$ (Fig.~\ref{curvature}b, left) and the second at $a\approx8.4$ (Fig.~\ref{curvature}b, right). Note that the surfaces chosen here are commensurate with the 4-fold symmetry of the original square lattice and we don't include randomness. After each transformation the lattice still preserves the 4-fold symmetry. Curvature-induced transitions might be seen in systems with a substrate of non-zero Poisson ratio, as a region of non-zero Gaussian curvature is likely to form upon stretching the substrate along the axial direction.

\subsection{Impurity-induced lattice transformations on a spiral surface}
\begin{figure}
\centering 
\includegraphics[width=3.4in]{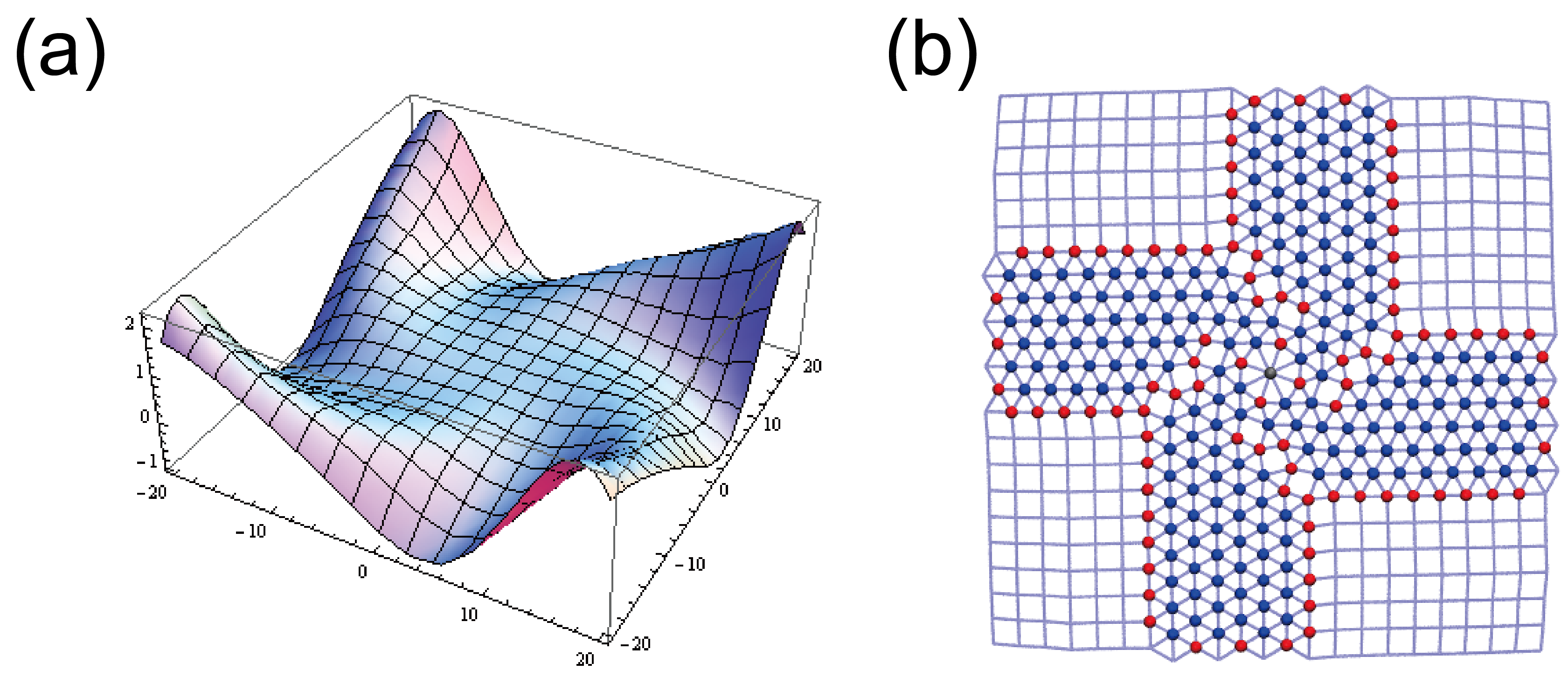} 
\caption{(Color online) (a) The spiral surface given by Eq.~[7]. (b) Connectivity (boundary not shown) at $t=10$, with the initial configuration same as in Fig.~1a but on the spiral surface.} 
\label{spiral}
\end{figure}

We next explored the lattice evolution with both impurities and spatial curvature. We position the system described in Fig.~\ref{square}a on a spiral surface (Fig.~\ref{spiral}a) given by 
\begin{eqnarray}
z= [(\frac{x}{a})^{2}+(\frac{y}{a})^{2}]^{\frac{3}{2}} \sin [4 \arctan (\frac{y}{x})+ 4 \sqrt{(\frac{x}{a})^{2}+(\frac{y}{a})^{2}}], \label{eq_spiral}
\end{eqnarray}
with $a=20$, and equilibrate it without exchange of water. Figure~\ref{spiral}b (movie \ref{spiral}b) shows the connectivity at $t=10$ after the exchange of water is turned on. The configuration exhibits a rotation of domains compared to its flat analogue. This is a simple example showing how curvature may affect the propagation of the rearrangement induced by one impurity; other patterns may be obtained by various curvature designs and impurity arrangements.

\section{Conclusion}
We have studied a 2D droplet system with an interaction described by a spring potential with a damping term. To explore the potential possibility of using the droplet network to study lattice transitions, we examine the transformation from a simple metastable state, a square lattice, to a triangular lattice, induced by either impurities or spatial curvature. We find that the subsequent propagation of rearrangements depends on the spring constant and the damping coefficient. When randomness is included, it may destroy the square lattice even without introducing additional defects if its strength is strong enough. Moreover, we show examples that different propagation patterns can be obtained through various impurity arrangements or curvature structures: the competition between transformation fronts induced by two impurity seeds leads to the formation of a domain wall; the  rearrangement takes place gradually from edge to central areas if we adapt an elliptic paraboloid or a hyperbolic paraboloid surface and increase its curvature slowly; an impurity-induced transformation on a spiral surface shows a rotation of domains.

\acknowledgments
DW thanks Meng Xiao for helpful discussions. We thank Syracuse University HTC Campus Grid and NSF award ACI-1341006 for providing computational resources. This research was supported by the Soft Matter Program of Syracuse University.

\section{Appendix A. Absorptive boundary conditions}
In the case of impurity-induced transformations, we add to the system a six-layer boundary with gradually increased damping values. For a system with a certain damping $\zeta$, the damping values of the six boundary layers (from innermost to outermost) are $2\zeta$, $4\zeta$, $10\zeta$, $10\zeta$, $20\zeta$, $20\zeta$. Thus for systems with $\zeta=10$ (Fig.~\ref{square}, c and f), the $\zeta$ values of the six boundary layers (from innermost to outermost) are 20, 40, 100, 100, 200, 200. For systems with $\zeta=1$ (Fig.~\ref{square}, d and g),  
the values are 2, 4, 10, 10, 20, 20, as plotted in Fig.~\ref{square}b. For systems with $\zeta=0.1$ (Fig.~\ref{square}, e and h), the values are 0.2, 0.4, 1.0, 1.0, 2.0, 2.0. For systems with $\zeta=0.2$ (Fig.~\ref{rectangular}, Fig.~\ref{random}, Fig.~\ref{curvature} and Fig.~\ref{spiral}), the values are 0.4, 0.8, 2.0, 2.0, 4.0, 4.0.

\section{Appendix B. Supplementary movies}
To better illustrate the transformations, we attach Movie \ref{square} (c to h), Movie \ref{rectangular} (c and d), Movie \ref{curvature} (a and b) and Movie \ref{spiral}, which match the number of figures, to the article. In the movies, color represent osmolarity, with high osmolarity $5/V_{0}$ red and low osmolarity $1/V_{0}$ blue. The gray droplets in Movie \ref{square} represent the absorptive boundary.


\bibliographystyle{epj}
\bibliography{droplets}

\begin{thebibliography}{35}

\bibitem{Tsonis2007}
D.M. McClung, in \emph{Nonlinear Dynamics in Geosciences}, edited by A.A.
  Tsonis, J.B. Elsner (Springer, New York, USA, 2007), chap.~24

\bibitem{Alberts2002}
B.~Alberts, A.~Johnson, J.~Lewis, M.~Raff, K.~Roberts, P.~Walter,
  \emph{Molecular Biology of the Cell, Fourth Edition} (Garland Science, New
  York, USA, 2002), chap.~3

\bibitem{Vogel1994}
W.~Vogel, \emph{Glass Chemistry, 2nd ed} (Springer, New York, 1994)

\bibitem{Moore2011}
E.B. Moore, V.~Molinero, Nature \textbf{479}, 506 (2011)

\bibitem{Russel1992}
W.B. Russel, D.A. Saville, W.R. Schowalter, \emph{Colloidal Dispersions}
  (Cambridge University Press, Cambridge, UK, 1992)

\bibitem{Bundy1980}
F.P. Bundy, J. Geophys. Res. \textbf{85}, 6930 (1980)

\bibitem{Porter1992}
D.A. Porter, K.E. Easterling, \emph{Phase Transformations in Metals and Alloys,
  Second Edition} (Chapman and Hall, London, UK, 1992)

\bibitem{Ise2005}
N.~Ise, I.~Sogami, \emph{Structure Formation in Solution} (Springer Berlin
  Heidelberg, Germany, 2005)

\bibitem{Gasser2009}
U.~Gasser, J. Phys.: Condens. Matter \textbf{21}, 203101 (2009)

\bibitem{Zhou2011}
H.~Zhou, S.~Xu, Z.~Sun, X.~Du, L.~Liu, Langmuir \textbf{27}, 7439 (2011)

\bibitem{Bayley2013}
G.~Villar, A.D. Graham, H.~Bayley, Science \textbf{340}, 48 (2013)

\bibitem{Princen1980}
M.P. Aronson, H.M. Princen, Nature \textbf{286}, 370 (1980)

\bibitem{Bayley2011}
G.~Villar, A.J. Heron, H.~Bayley, Nat. Nanotechnol. \textbf{6}, 803 (2011)

\bibitem{Tao2015}
T.~Zhang, D.~Wan, J.M. Schwarz, M.J. Bowick, arXiv:1505.01108v2

\bibitem{Han2015}
Y.~Peng, F.~Wang, Z.~Wang, A.M. Alsayed, Z.~Zhang, A.G. Yodh, Y.~Han, Nat.
  Mater. \textbf{14}, 101 (2015)

\bibitem{Lekkerkerker2005}
V.W.A. de~Villeneuve, R.P.A. Dullens, D.G.A.L. Aarts, E.~Groeneveld, J.H.
  Scherff, W.K. Kegel, H.N.W. Lekkerkerker, Science \textbf{309}, 1231 (2005)

\bibitem{Menon2002}
K.~Feitosa, N.~Menon, Phys. Rev. Lett. \textbf{88}, 198301 (2002)

\bibitem{Yao2014}
Z.~Yao, M.O. de~la Cruz, Proc. Natl. Acad. Sci. \textbf{111}, 5094 (2014)

\bibitem{Thomson1904}
J.J. Thomson, Phil. Mag. \textbf{7}, 237 (1904)

\bibitem{Yao2013}
Z.~Yao, M.O. de~la Cruz, Phys. Rev. Lett. \textbf{111}, 115503 (2013)

\bibitem{Funkhouser2013}
C.M. Funkhouser, R.~Sknepnek, M.O. de~la Cruz, Soft Matter \textbf{9}, 60
  (2013)

\bibitem{Bausch2003}
A.R. Bausch, M.J. Bowick, A.~Cacciuto, A.D. Dinsmore, M.F. Hsu, D.R. Nelson,
  M.G. Nikolaides, A.~Travesset, D.A. Weitz, Science \textbf{299}, 1716 (2003)

\bibitem{Irvine2010}
W.T.M. Irvine, V.~Vitelli, P.M. Chaikin, Nature \textbf{468}, 947 (2010)

\bibitem{Irvine2012}
W.T.M. Irvine, M.J. Bowick, P.M. Chaikin, Nat. Mater. \textbf{11}, 948 (2012)

\bibitem{Bowick2011}
M.J. Bowick, Z.~Yao, Europhys. Lett. \textbf{93}, 36001 (2011)

\bibitem{Lidmar2003}
J.~Lidmar, L.~Mirny, D.R. Nelson, Phys. Rev. E \textbf{68}, 051910 (2003)

\bibitem{Wan2015}
D.~Wan, M.J. Bowick, R.~Sknepnek, Phys. Rev. E \textbf{91}, 033205 (2015)

\bibitem{Humphrey1996}
W.~Humphrey, A.~Dalke, K.~Schulten, J. Mol. Graphics \textbf{14}, 33 (1996)

\bibitem{Stone1998}
J.~Stone, Master's thesis, University of Missouri-Rolla (1998)

\bibitem{Santos1995}
F.D.D. Santos, T.~Ondar\c{c}uhu, Phys. Rev. Lett. \textbf{75}, 2972 (1995)

\bibitem{Lee2000}
S.W. Lee, P.E. Laibinis, J. Am. Chem. Soc. \textbf{122}, 5395 (2000)

\bibitem{Sumino2005}
Y.~Sumino, N.~Magome, T.~Hamada, K.~Yoshikawa, Phys. Rev. Lett. \textbf{94},
  068301 (2005)

\bibitem{John2005}
K.~John, M.~B\"{a}r, U.~Thiele, Eur. Phys. J. E \textbf{18}, 183 (2005)

\bibitem{Decicco1947}
J.~DeCicco, Bull. Amer. Math. Soc. \textbf{53}, 993 (1947)

\bibitem{Kreyszig1991}
E.~Kreyszig, \emph{Differential Geometry} (Dover, New York, 1991)

\end{thebibliography}

\end{document}